\documentclass[english,times,superscriptaddress]{revtex4-1}
\usepackage[T1]{fontenc}
\usepackage[latin9]{inputenc}
\usepackage{amsmath}
\usepackage{amssymb}
\usepackage{graphicx}

\makeatletter
\usepackage{times}

\makeatother

\usepackage{babel}
\begin{document}
\title{Synchronization of two cavity-coupled qubits measured by entanglement}
\author{Tian-tian Huan}
\address{Institute of Applied Physics and Materials Engineering, University
of Macau, Macau, China}
\author{Ri-gui Zhou}
\affiliation{\textbackslash address\{College of Information Engineering, Shanghai
Maritime University, Shanghai 201306, China}
\author{Hou Ian}
\email{houian@um.edu.mo}

\address{Institute of Applied Physics and Materials Engineering, University
of Macau, Macau, China}
\affiliation{Zhuhai UM Science \& Technology Research Institute, Zhuhai, Guangdong,
China}
\begin{abstract}
Some nonlinear radiations such as superfluorescence can be understood
as cooperative effects between atoms. We regard the cooperative radiation
as a manifested effect secondary to the intrinsic synchronization
among the atoms and propose a time-resolved measure of synchronization
on a cavity-coupled dual-qubit system using multipartite concurrence.
Comparing the variation of the concurrence over time with that of
an asynchronicity measure, we find that the synchronization between
the qubits features a time delay characteristic to the initiation
of superfluorescent pulses. The delay coincides with the duration
for the qubits to establish cooperation and emit the collective radiation,
after which the concurrence monotonically increases to a stationary
value while the asynchronicity dives to a steady minimum. Furthermore,
the establishment of synchronization is determined by the qubit-cavity
coupling strength. Asynchronicity shows that synchronization is only
possible when the coupling enters strong regime and sustains to a
high level when the coupling becomes ultra-strong.
\end{abstract}
\maketitle

\section{Introduction}

Centuries ago, Huygens studied the correlation among the motions of
pendulums and discovered the synchronization pattern of these individual
oscillators under the influence of a common oscillator they are coupled
to~\citep{PIKOVSKY}. How synchronizations arise in different situations
has since remained a problem of interest~\citep{acebron05}. In recent
years, the study of the classical phenomenon has been revived under
the quantum regime. Synchronization is observed between a pair of
nanomechanical oscillators~\citep{shim07} and on the motions of
the Cooper pairs among Josephson-insulated superconducting islands~\citep{vinokur08}.
It is also ubiquitously predicted between a qubit and an oscillator~\citep{zhirov08},
between a cavity field and an oscillator~\citep{ying14}, between
two oscillators~\citep{qiao18}, among a trapped group of cold atoms~\citep{heimonen18},
and even between a quantum Van der Pol oscillator and an external
drive~\citep{walter18}.

Here we study the synchronization of two qubits coupled indirectly
to each other through a cavity field, in specific relevance to the
quantum optical phenomenon of superfluorescence~\citep{bonifacio75}.
This fluorescent effect embodies Dicke's formulation of superradiance~\citep{dicke54},
under which the radiation intensity of the emitted photons from an
$N$-atom ensemble is proportional to $N^{2}$, in the observable
time domain. Superfluorescence theory shows that the emission of the
$N^{2}$-dependent superradiance exhibits a positive time delay~\citep{haake79,polder79}.
Experimentally recorded on hydrofluoric gas~\citep{skriban73}, cesium~\citep{gibbs77},
and most recently rubidium vapor~\citep{ariun10}, this delay shows
the necessity of a finite time duration which the atoms use to establish
cooperation before the radiation is initiated~\citep{arecchi70}.

Such a delay is also characteristic of synchronization: it takes finite
time for the Huygens pendulums to undergo oscillations in-phase, which
is also registered for quantum synchronization processes~\citep{yokoshi17,greilich06}.
We show in this paper that two cavity-coupled qubits would exhibit
a time delay while undergoing a dynamic entangling process. Similar
to the entanglement measures registered on other quantum systems~\citep{abdi12,tian13,huan15},
the tripartite qubit-cavity-qubit system studied here demonstrates
characteristic time-dependent degrees of entanglement. Measured in
both bipartite and tripartite concurrences~\citep{wootters98,coffman00,rungta01}
which is appropriate for the Hilbert space of the discrete-level system
whose dimension satisfies the requirement to produce synchronization~\citep{roulet18},
the entanglement is shown universally to begin at a zero level and
rise to a saturated value for a finite duration of time. The saturated
entangled state is sustained thereafter, demonstrating the features
of a synchronized state.

Concurrence is proved to be a good measure for static analysis of
collective phenomena such as radiation~\citep{ian16,ian14}. To verify
the coincidence between the maximization of concurrence and the dynamic
process of synchronization, we introduce a synchronization measure
computed from the density matrix, which is modified from the synchronization
measure introduced on the $(x,p)$-quadratures of quantum oscillators~\citep{mari13}.
The transition point in time produced from the synchronization measures
matches exactly the delay time found in the concurrence evolution,
proving that the qubit-to-qubit synchronization is well registered
in the entanglement. Moreover, since two-qubit systems on a superconducting
circuit can produce superfluorescent pulses~\citep{mlynek14}, the
synchronization delay is associated with the initiation of superradiance,
thereby establishing the dynamic correlation between entanglement
and the atomic cooperation for collective phenomena.

\section{Results\label{sec:results}}

\subsection{The tripartite system}

\begin{figure}
\includegraphics[bb=0bp 0bp 705bp 320bp,clip,width=8.8cm]{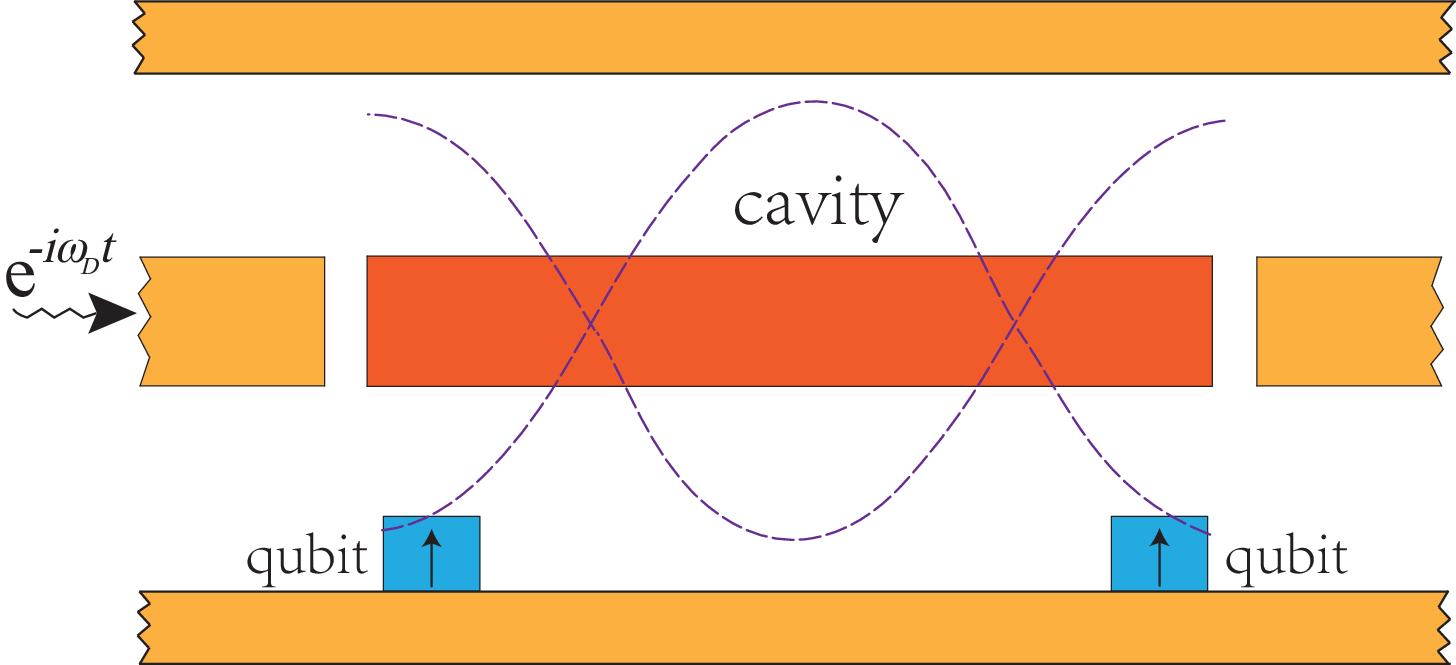}

\caption{Illustration of the tripartite system: two superconducting qubits
is coupled to the cavity field (indicated by the red strip), through
which entanglement between the qubits are generated over time. The
cavity field is driven by an incident field entered from the left.~\label{fig:model}}
\end{figure}

The derivation for the dynamics discussed above is modeled on a generic
cavity quantum electrodynamic (QED) system where each qubit is coupled
to the cavity through dipole-field interaction under the rotating
wave approximation. The parameters adopted for the numerical analysis
are sourced from the superconducting circuit implementation~\citep{blais04,majer07}
of the cavity QED system. Hence, the tripartite system can be illustrated
from the model figure~\ref{fig:model}, where the cavity is indicated
by the stripped waveguide and the qubits are located at the anti-nodes
of the cavity field to ensure maximum coupling.

The total system Hamiltonian $H=H_{0}+H_{\text{\ensuremath{\mathrm{int}}}}+H_{\mathrm{ext}}$
is composed of three parts: the free Hamiltonian, the interactions
among the system components, and the external driving, which reads,
respectively, $(\hbar=1)$
\begin{align}
H_{0} & =\omega_{c}a^{\dagger}a+\Omega_{L}\sigma_{L,z}+\Omega_{R}\sigma_{R,z},\label{eq:free system}\\
H_{\mathrm{int}} & =\eta_{L}\left(a\sigma_{L,+}+a^{\dagger}\sigma_{L,-}\right)+\eta_{R}\left(a\sigma_{R,+}+a^{\dagger}\sigma_{R,-}\right),\label{eq:interaction}\\
H_{\mathrm{ext}} & =i\varepsilon_{D}\left(a^{\dagger}e^{-i\omega_{D}t}-ae^{i\omega_{D}t}\right).\label{eq:external Hamiltonian}
\end{align}
In $H_{0}$, $\omega_{c}$ denotes the frequency of the cavity mode
and $\Omega_{L}$ ($\Omega_{R}$) denotes the transition frequency
of the left (right) qubit, associated with the Pauli matrix $\sigma_{L,z}$
($\sigma_{R,z}$). In $H_{\mathrm{int}}$, $\eta_{L}(\eta_{R})$ denotes
the coupling strength to the left (right) qubit. In $H_{\mathrm{ext}}$,
the external driving field has frequency $\omega_{D}$ and driving
strength $\varepsilon_{D}$. 

The combined system of a cavity and two qubits has its bare states
described by the tensor product state $\{|e_{L}\rangle,|g_{L}\rangle\}\otimes\{|n\rangle\}\otimes\{|e_{R}\rangle,|g_{R}\rangle\}$,
where $|e_{L}\rangle$ ($|g_{L}\rangle$) denotes the excited (ground)
state of the left (right) qubit; $|n\rangle$ denotes the Fock number
states of the cavity mode; $|e_{R}\rangle$ ($|g_{R}\rangle$) denotes
the excited (ground) state of the right qubit. To simplifying the
notation, we omit the subscripts $L$ and $R$ when writing the product
states and let the first letter denote the state of the left qubit,
the middle letter that of the cavity mode, and the last letter that
of the right qubit (e.g. $|e,n,g\rangle=|e_{L}\rangle\otimes|n\rangle\otimes|g_{R}\rangle$).

The free Hamiltonian $H_{0}$ and the interaction Hamiltonian $H_{{\rm int}}$
constitute a closed subsystem, for which there exist dressed states
that diagonalize $H_{0}+H_{\mathrm{int}}$. To find an analytical
expression for the dressed states, we consider the sets of energy-conserving
states $|e,n,g\rangle$, $|g,n+1,g\rangle$, and $|g,n,e\rangle$,
which are resonant within single-photon processes, to contribute to
a dressed state for each $n$. In other words, the state $|e,n,e\rangle$
which is resonant with $|g,n+2,g\rangle$ through a double-photon
process is avoided. These single-photon resonant states form an invariant
subspace, for which the closed Hamiltonian consists of $3\times3$
symmetric block matrices. Therefore, we have the eigen-equation

\begin{eqnarray}
\left(H_{\mathrm{0}}+H_{{\rm int}}\right)|u_{k}^{(n)}\rangle & = & E_{k}^{(n)}\left|u_{k}^{(n)}\right\rangle ,\label{eq:eigenequation}
\end{eqnarray}
where the eigenvectors $|u_{k}^{(n)}\rangle$ denote the dressed states
that diagonalize the $3n\times3n$ matrix $H_{0}+H_{\mathrm{int}}$
and the eigenvalues $E_{k}^{(n)}$ denote the dressed-state energies
in the diagonalized space. The index $k$ enumerates $\{1,2,3\}$
to indicate the dressed levels within the $n$-th cluster. 

Block-diagonalizing $H_{0}+H_{\mathrm{int}}$ for Eq.~(\ref{eq:eigenequation})
results in a cubic equation of $E_{k}^{(n)}$ for each $n$, whose
roots are
\begin{equation}
E_{k}^{\left(n\right)}=\frac{2}{3}\sqrt{\delta_{n}^{2}+3(\Delta^{2}+\eta_{L}^{2}+\eta_{R}^{2})}\cos\left(\theta+\frac{2k\pi}{3}\right)+n\omega_{c}+\frac{\delta_{n}}{3},\label{eq:eigenvalue}
\end{equation}

where the $\Delta=\Omega_{L}-\Omega_{R}$ denotes the left-right qubit
detuning, $\delta=\omega_{c}-\Omega_{L}-\Omega_{R}$ is the detuning
between the cavity and the two qubits, and the angle is defined as
\begin{equation}
\theta=\frac{1}{3}\cos^{-1}2\left(\delta^{2}+3\Delta^{2}+3(\eta_{L}^{2}+\eta_{R}^{2})\right)^{-3/2}\Bigl[2\delta^{3}+9\delta\left(\eta_{L}^{2}+\eta_{R}^{2}\right)-18\delta\Delta^{2}+27\Delta\left(\eta_{L}^{2}-\eta_{R}^{2}\right)\Bigr].\label{eq:theta}
\end{equation}
The corresponding eigenvector reads

\begin{equation}
|u_{k}^{(n)}\rangle=\alpha_{L,k}^{(n)}|e,n,g\rangle+\alpha_{C,k}^{(n)}|g,n+1,g\rangle+\alpha_{R,k}^{(n)}|g,n,e\rangle,\label{eq:dressed_state}
\end{equation}
where the transformation coefficients are 
\begin{align}
\alpha_{L,k}^{(n)} & =-\eta_{L}(\Delta-n\omega_{c}+E_{k}^{(n)})\Bigl/Z_{k}^{(n)},\label{eq:alpha_L}\\
\alpha_{C,k}^{(n)} & =\Bigl[\Delta^{2}-(E_{k}^{(n)}-n\omega_{c})^{2}\Bigr]\Bigl/Z_{k}^{(n)},\label{eq:alpha_C}\\
\alpha_{R,k}^{(n)} & =\eta_{R}(\Delta-E_{k}^{(n)}+n\omega_{c})\Bigl/Z_{k}^{(n)},\label{eq:alpha_R}
\end{align}
with $Z_{k}^{(n)}$ being the normalization constant
\begin{equation}
Z_{k}^{(n)}=\left(\eta_{L}^{2}\left[\Delta+E_{k}^{(n)}-n\omega_{c}\right]^{2}+\eta_{R}^{2}\Bigl[\Delta-E_{k}^{(n)}+n\omega_{c}\Bigr]^{2}+\left[\Delta^{2}-(E_{k}^{(n)}-n\omega_{c})^{2}\right]^{2}\right)^{1/2}.
\end{equation}
The detailed derivation is given in the Methods section.

In the dressed space spanned by the basis vectors of Eq.~(\ref{eq:dressed_state}),
the closed Hamiltonian is written in the diagonalized form 
\begin{eqnarray}
H_{\mathrm{0}}+H_{{\rm int}} & = & \sum_{n,k}E_{k}^{(n)}|u_{k}^{(n)}\rangle\langle u_{k}^{(n)}|,
\end{eqnarray}
while the annihilation operator 
\begin{align}
a & =\mathbb{I}_{\mathrm{L}}\otimes a\otimes\mathbb{I}_{\mathrm{R}}\nonumber \\
 & \approx\sum_{n}|g,n,e\rangle\langle g,n+1,e|+|e,n,g\rangle\langle e,n+1,g|+\sqrt{2}|g,n+1,g\rangle\langle g,n+2,g|
\end{align}
under the single-photon processes is transformed to
\begin{equation}
a=\sum_{n,j,k}\Bigl[\alpha_{L,j}^{(n)*}\alpha_{L,k}^{(n+1)}+\sqrt{2}\alpha_{C,j}^{(n)*}\alpha_{C,k}^{(n+1)}+\alpha_{R,j}^{(n)*}\alpha_{R,k}^{(n+1)}\Bigl]|u_{j}^{(n)}\rangle\langle u_{k}^{(n+1)}|\label{eq:a_dressed}
\end{equation}
where the indices $j$ and $k$ enumerate over the set $\{1,2,3\}$.

The permitted dressed level transitions induced by the external driving
can be found by substituting Eq.~(\ref{eq:a_dressed}) into Eq.~(\ref{eq:external Hamiltonian}).
In the weak-energy limit where the transitions are confined to the
lowest two clusters of states ($n=0$ and $n=1$), the total Hamiltonian
is written as
\begin{align}
H^{\left(0,1\right)}= & \sum_{j}\Bigl[E_{j}^{(0)}|u_{j}^{(0)}\rangle\langle u_{j}^{(0)}|+E_{j}^{(1)}|u_{j}^{(1)}\rangle\langle u_{j}^{(1)}|\Bigr]-\sum_{j,k}i\varepsilon_{D}e^{i\omega_{D}t}\nonumber \\
 & \Bigl[\Bigl(\alpha_{L,j}^{(0)*}\alpha_{L,k}^{(1)}+\sqrt{2}\alpha_{C,j}^{(0)*}\alpha_{C,k}^{(1)}+\alpha_{R,j}^{(0)*}\alpha_{R,k}^{(1)}\Bigr)|u_{j}^{(0)}\rangle\langle u_{k}^{(1)}|+\mathrm{H.c.}\Bigl]\label{eq:H_dressed}
\end{align}
in the dressed space. Introducing the time-dependent state vector

\begin{eqnarray}
\left|\psi(t)\right\rangle  & = & \sum_{j}\left(c_{j}(t)|u_{j}^{(0)}\rangle+d_{j}(t)|u_{j}^{(1)}\rangle\right)\label{eq:sys_state_dressed}
\end{eqnarray}
in the confined state space and applying it to the Hamiltonian above,
one has the Schr\"{o}dinger equations of the time coefficients

\begin{eqnarray}
\dot{c}_{j}(t) & = & -iE_{j}^{(0)}c_{j}(t)-\varepsilon_{D}e^{i\omega_{D}t}\lambda_{l}\alpha_{l,j}^{(0)*}\alpha_{l,k}^{(1)}d_{k}(t),\\
\dot{d}_{j}(t) & = & -iE_{j}^{(1)}d_{j}(t)+\varepsilon_{D}e^{-i\omega_{D}t}\lambda_{l}\alpha_{l,j}^{(1)*}\alpha_{l,k}^{(0)}c_{k}(t),
\end{eqnarray}
where $\lambda_{l}$ denotes the weight of the summation over the
index $l$ for the system component $L$ (the left qubit), $R$ (the
right qubit), or $C$ (the cavity), i.e. $\lambda_{L}=\lambda_{R}=1$
and $\lambda_{C}=\sqrt{2}$. In the equations, we observe the Einstein
summation convention.

In the rotating frame $c_{j}\left(t\right)=c_{j}^{\prime}\left(t\right)\mathrm{exp}\{-iE_{j}^{(0)}t\}$
and $d_{j}(t)=d_{j}^{\prime}(t)\mathrm{exp}\{-iE_{j}^{(1)}t\}$, the
coupled equations can be written as the linear homogeneous system
of differential equations $\dot{\mathbf{c}^{\prime}}=A\mathbf{c}^{\prime}$
where $\mathbf{c^{\prime}}=\left[c_{1}^{\prime}\;c_{2}^{\prime}\;c_{3}^{\prime}\;d_{1}^{\prime}\;d_{2}^{\prime}\;d_{3}^{\prime}\right]$
and

\begin{equation}
A=\left[\begin{array}{cc}
0 & -\left[\varepsilon_{D}e^{i\zeta_{kj}t}\lambda_{l}\alpha_{l,j}^{(0)*}\alpha_{l,k}^{(1)}\right]\\
\left[\varepsilon_{D}e^{-i\zeta_{jk}t}\lambda_{l}\alpha_{l,j}^{(1)*}\alpha_{l,k}^{(0)}\right] & 0
\end{array}\right]
\end{equation}
where the square brackets $[\cdot]$ indicate $3\times3$ submatrices
in the matrix $A$ with $j$ and $k$ being the row and the column
indices, respectively. We denote $\zeta_{kj}=\omega_{D}-\left(E_{k}^{\left(1\right)}-E_{j}^{\left(0\right)}\right)$
for the detuning between the driving and the dressed states. Since
$A$ is integrable, then solving the linear system for $\{c_{j},d_{j}\}$
and expanding the dressed states by using the bare states in Eq.~(\ref{eq:sys_state_dressed}),
one can find the expansion coefficients $\gamma$ of the state vector

\begin{multline}
|\psi(t)\rangle=\gamma_{L}^{(0)}\left(t\right)|e,0,g\rangle+\gamma_{C}^{(0)}\left(t\right)|g,1,g\rangle+\gamma_{R}^{(0)}\left(t\right)|g,0,e\rangle\\
+\gamma_{L}^{(1)}\left(t\right)|e,1,g\rangle+\gamma_{C}^{(1)}\left(t\right)|g,2,g\rangle+\gamma_{R}^{(1)}\left(t\right)|g,1,e\rangle\label{eq:sys_state}
\end{multline}
back in the bare state space.

\subsection{Evolution of the state vector}

To see that the evolution of the state vector can initiate the synchronization
of the two delocalized qubits, we assume the cavity mode is initially
driven by the external field to reach a partial population inversion
while setting the qubits initially at the ground. In other words,
the expansion coefficients at the initial moment are: $\gamma_{C}^{\left(0\right)}=\sqrt{0.9}$,
$\gamma_{C}^{\left(1\right)}=\sqrt{0.1}$, and $\gamma_{L}^{\left(0\right)}=\gamma_{R}^{\left(0\right)}=\gamma_{L}^{\left(1\right)}=\gamma_{R}^{\left(1\right)}=0$.

\begin{figure}
\includegraphics[bb=0bp 45bp 550bp 500bp,clip,width=8.8cm]{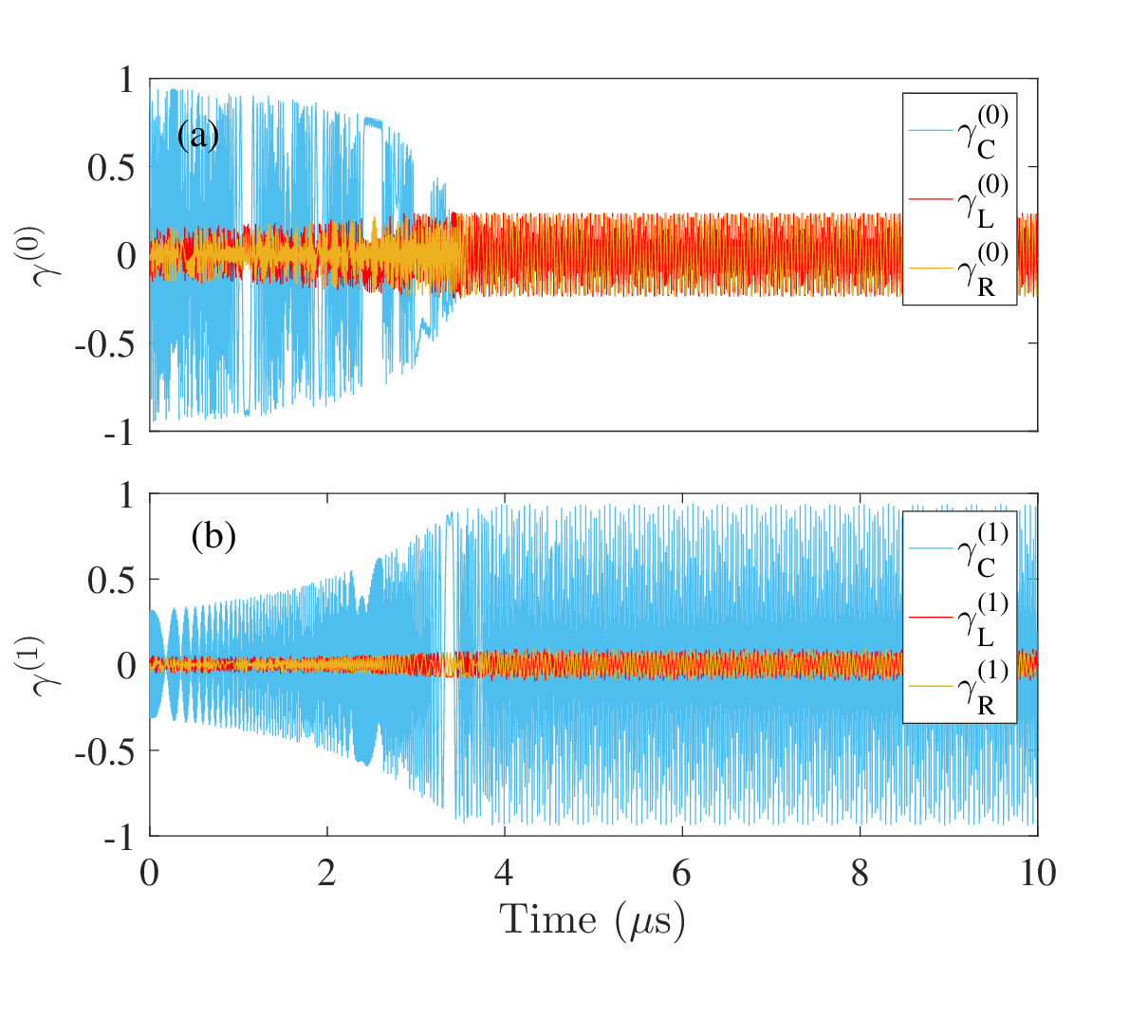}

\caption{The time evolutions of the six expansion coefficients: (a) for $n=0$;
(b) for $n=1$. The red, the blue, and the yellow curves associate
with the state $|e,n,g\rangle$, $|g,n+1,g\rangle$, and $|g,n,e\rangle$,
respectively.~\label{fig:coefficient}}
\end{figure}

We plot out the evolutions of these coefficients in Fig.~\ref{fig:coefficient},
using the experimentally accessible parameters of superconducting
charge-phase qubits~\citep{majer07}: $\Omega_{L}/2\pi=\Omega_{R}/2\pi=6.1$~GHz,
$\eta_{L}/2\pi=\eta_{R}/2\pi=500$~MHz, and $\omega_{c}/2\pi=6.32$~GHz.
The frequency of the external field is maintained at $\omega_{D}/2\pi=5.3$~GHz.
The lower set of states with $n=0$ is given in Fig.~\ref{fig:coefficient}(a)
whereas the upper set with $n=1$ is given in Fig.~\ref{fig:coefficient}(b).

We observe that for both the lower set and the upper set of states,
there exists a transition point of the oscillations of the coefficients,
which is located at about $3.7\mu\mathrm{s}$ in the plots. In particular,
$\gamma_{C}^{\left(0\right)}$ is transited from a region of shrinking
oscillation to a region of small fluctuation at this point. Meanwhile,
$\gamma_{L}^{\left(1\right)}$,$\gamma_{R}^{\left(1\right)}$, and
$\gamma_{C}^{\left(1\right)}$ are transited from an amplifying region
to a region of saturated oscillation envelope. The contrasting behavior
of the two sets of coefficients demonstrates that the energy excitation
that exists in the cavity mode is transferred to the left and the
right qubits whose complementary oscillations imply the build-up of
the entanglement between them. 

\subsection{Bipartite and tripartite concurrences}

To fully capture the evolution characteristics of the two cavity-coupled
qubits from a holistic point of view, we apply two entanglement measures
-- bipartite concurrence and tripartite concurrence -- to the state
vector of the total system.

The bipartite concurrence quantifies the inseparability of the joint
pure state of two coupled systems of arbitrary dimensions by inverting
the density matrix. For our case here, the joint state is the product
state $|\psi_{LR}\rangle$ of the indirectly coupled left and right
qubits. Thus the inversion is conducted through the superoperator
$S_{D_{1}}\otimes S_{D_{2}}$ where the dimensions $D_{1}=D_{2}=2$
and the bipartite concurrence is defined as $C_{2}(\psi_{LR})=\sqrt{\langle\psi_{LR}|S_{2}\otimes S_{2}\left(|\psi_{LR}\rangle\langle\psi_{LR}|\right)|\psi_{LR}\rangle}$.
Given the consideration of pure states, for which $\mathrm{tr}\rho^{2}=1$
and $\mathrm{tr}\rho_{L}^{2}=\mathrm{tr}\rho_{R}^{2}$, the definition
reduces to $C_{2}(\psi_{LR})=\sqrt{2\left[1-\mathrm{tr}\left(\rho_{L}^{2}\right)\right]}$
where $\rho_{L}=\mathrm{tr}_{R}\left(\mathrm{tr}_{C}\left(|\psi\rangle\langle\psi|\right)\right)$
is the reduced density matrix of the left qubit. 

\begin{figure}
\includegraphics[bb=0bp 0bp 550bp 500bp,clip,width=8.8cm]{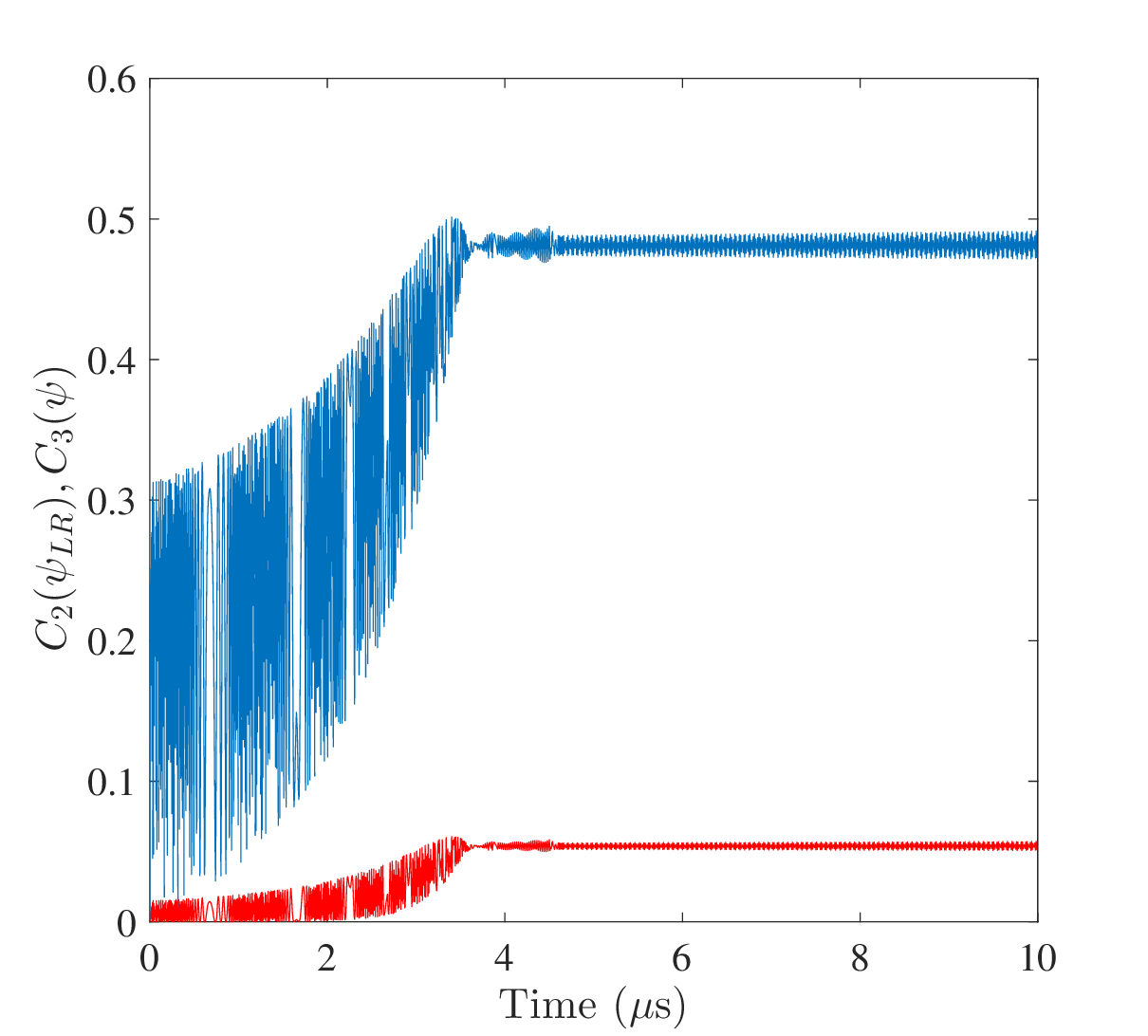}\caption{Time evolution of the bipartite concurrence $C_{2}\left(\psi_{LR}\right)$
between the two qubits (blue) and the tripartite concurrence $C_{3}\left(\psi\right)$
among the qubits and the cavity (red). A symmetric scenario is assumed
between the left and the right qubits: $\eta_{L}/2\pi=\eta_{R}/2\pi=500$~MHz
and $\Omega_{L}/2\pi=\Omega_{R}/2\pi=6.1$~GHz.~\label{fig:2-con.}}
\end{figure}

Applying $|\psi\left(t\right)\rangle$ in Eq.~(\ref{eq:sys_state})
to the formula, we derive the evolution of the bipartite concurrence,
shown as the blue curve in Fig.~\ref{fig:2-con.}. It becomes apparent
that the transition point that manifests in Figs.~\ref{fig:coefficient}(a)
and \ref{fig:coefficient}(b) signifies the concurrence reaching a
maximum after a gradual monotonic increase in the oscillation envelope.
This maximum concurrence is retained thereafter. The finite delay
time $\tau_{D}=3.7\mu\mathrm{s}$ that the concurrence spends to reach
its maximal value reflects the time the two qubits use to reach a
maximal synchronization through their mutual couplings to the cavity
mode.

The cavity mode plays an active part in initiating the entanglement
between the two qubits. From the entanglement-theoretic point of view,
the concurrence is distributed among the qubits as well as the cavity.
Taking away the pairwise entanglements between any two parties in
the tripartite system, one obtains the residual concurrence that remains
as an equally distributed entanglement among all three parties~\citep{coffman00}.
Extending the original formulation on three-qubit systems, we generalize
the inversion operations for two arbitrary-dimensional systems given
above to three arbitrary-dimensional system. That is, we introduce
the superoperator
\begin{align}
S_{D_{1}}\otimes S_{D_{2}}\otimes S_{D_{3}}\left(\rho\right)= & I\otimes I\otimes I-I\otimes I\otimes\rho_{R}-\rho_{L}\otimes I\otimes I-I\otimes\rho_{C}\otimes I\nonumber \\
 & +\rho_{LR}\otimes I+I\otimes\rho_{CR}+\rho_{LC}\otimes I-\rho.\label{eq:superoperator}
\end{align}
for our $D_{1}\times D_{2}\times D_{3}$ dimension tripartite system,
where $D_{1}=D_{3}=2$ for the qubits and $D_{2}=n$ for the cavity
mode. In Eq.~(\ref{eq:superoperator}), $I$ denotes the identity
matrix while $\rho_{L}$, $\rho_{C}$, $\rho_{R}$, $\rho_{LR}$,
$\rho_{CR}$, and $\rho_{LC}$ denote the reduced density matrices
of the components and the two-component subsystems. Applying the inversion,
we thus derive a tripartite residual concurrence
\begin{align}
C_{3}\left(\psi\right) & =\sqrt{\langle\psi|S_{D_{1}}\otimes S_{D_{2}}\otimes S_{D_{3}}\left(|\psi\rangle\langle\psi|\right)|\psi\rangle}\nonumber \\
 & =\sqrt{1-\mathrm{tr}\rho_{R}^{2}-\mathrm{tr}\rho_{L}^{2}-\mathrm{tr}\rho_{C}^{2}+\mathrm{tr}\rho_{LR}^{2}+\mathrm{tr}\rho_{CR}^{2}+\mathrm{tr}\rho_{LC}^{2}-\mathrm{tr}\rho^{2}}.
\end{align}

Again, using Eq.~(\ref{eq:sys_state}), we plot the tripartite concurrence
as the red curve in Fig.~\ref{fig:2-con.}. One can verify from the
plot that the residual concurrence evolves in a similar fashion, which
contains a signifying transition point at the exactly same location
$\tau_{D}$ as that of the bipartite concurrence. Before $\tau_{D}$,
it arises from a zero value under a similarly increasing envelope
whereas, after $\tau_{D}$, it retains a non-zero saturated value.
The identical delay time again demonstrates the duration that the
system components spend on cooperation before maximal synchronization
is reached.

Comparing Fig.~\ref{fig:coefficient} and Fig.~\ref{fig:2-con.},
one sees that the energy quantum first dwells on the cavity mode ($|g,1,g\rangle$
and $|g,2,g\rangle$) without being emitted and absorbed by the qubits.
Only when the two qubits start to establish a cooperated motion does
the qubit-cavity-qubit resonance become effective such that the qubits
be excited to their respective excited states $|e,1,g\rangle$ and
$|g,1,e\rangle$. The entanglement is also established among the three
components when the excitation commences.

\begin{figure}
\includegraphics[bb=0bp 0bp 550bp 500bp,clip,width=8.8cm]{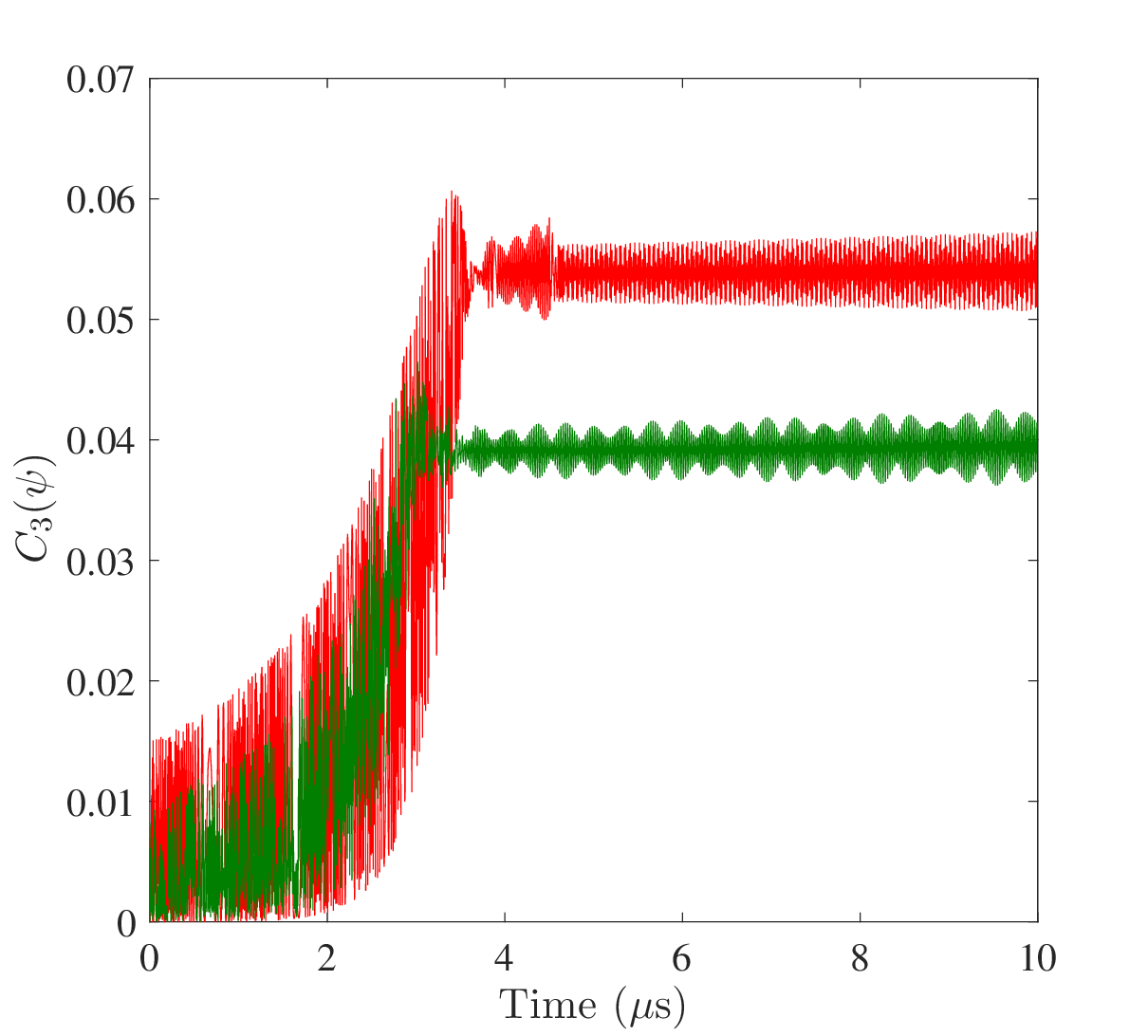}\caption{Comparison of the time evolutions of the tripartite concurrences between
the symmetric (red curve) and the asymmetric (green curve) scenarios.
For the symmetric case, both qubits are set to $\Omega_{L}/2\pi=\Omega_{R}/2\pi=6.1~\mathrm{GHz}$.
For the asymmetric case, the right qubit is adjusted to $\Omega_{R}/2\pi=7.1$GHz.
Coupling strengths are retained at $\eta_{L}/2\pi=\eta_{R}/2\pi=500$MHz
throughout.\label{fig:3-con.}}
\end{figure}

The concurrences plotted in Fig.~\ref{fig:2-con.} are computed upon
a symmetric setting of system parameters: the level spacings and the
coupling strengths of the qubits are assumed identical. The tripartite
concurrence of an asymmetric scenario with the right qubit level spacing
raised to $\Omega_{R}/2\pi=7.1\,\mathrm{GHz}$ is shown as the green
curve in Fig.~\ref{fig:3-con.} while the rest of the parameters
remain unchanged. For comparison, the symmetric case is plotted as
the red curve in the background. We observe that the delay to saturated
synchronization is inversely correlated with the larger eigenfrequency
out of the two qubits. For the case in Fig.~\ref{fig:3-con.}, increasing
$\Omega_{R}$ reduces delay time $\tau_{D}$. On the other hand, symmetric
settings lead to maximal synchronization at saturation. The asymmetric
case given by the green curve has the saturated synchronization reduced
to a lower level. Simulation under parameters set to various (not
shown in figures) verify these observations. 

\begin{figure}
\includegraphics[bb=0bp 0bp 440bp 320bp,clip,width=8.8cm]{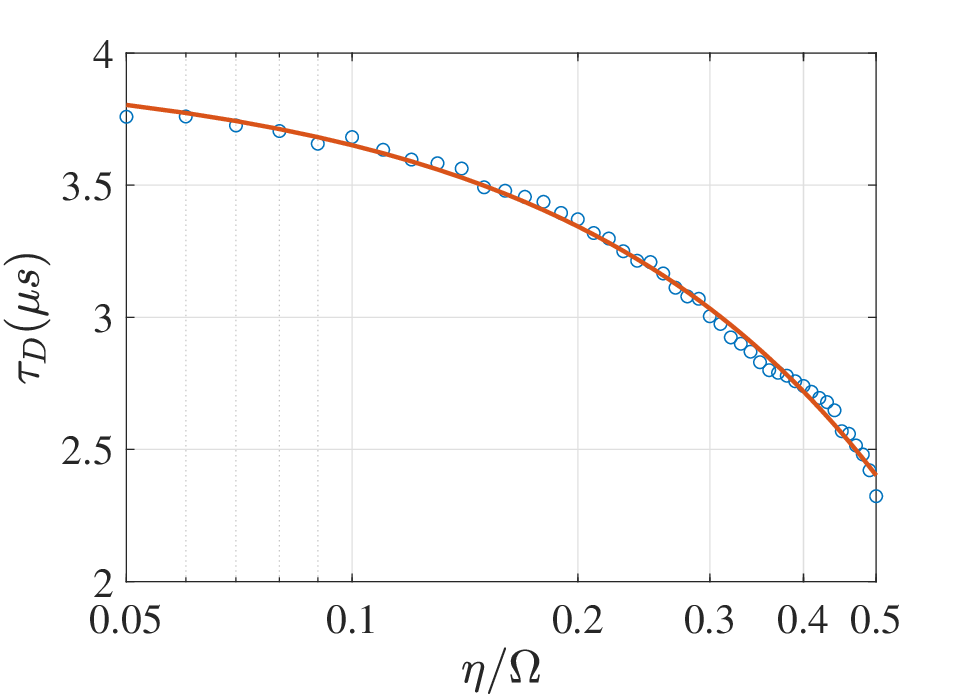}

\caption{The time delay $\tau_{D}$ is plotted as a function of the dimensionless
coupling strength $\eta/\Omega$ in a semilog scale. Symmetric setting
is assumed: the qubit level spacings $\Omega/2\pi=\Omega_{L,R}/2\pi=6.1$~GHz
and the coupling strength $\eta=\eta_{L,R}$. The circles indicate
the data points in the simulation runnings.~\label{fig:time delay}}
\end{figure}

The synchronization between the qubits is also affected by how strong
they are driven by the cavity mode, i.e. the coupling strengths $\eta_{L}$
and $\eta_{R}$. Shown in Fig.~\ref{fig:time delay} for the symmetric
scenario $\eta_{L}=\eta_{R}$ in a semilog plot, the greater is the
coupling, the lesser is the delay $\tau_{D}$. The numerical fit shows
that the delay obeys a quadratic relation over the exponential increase
in coupling strength. After the delay, a shorter delay time is associated
with a higher saturated level of concurrence, showing a stronger synchronization
between the qubits are reflected in both short delay and higher entanglement
measure.

\subsection{Asynchronicity}

Multi-partite concurrence as a measure of synchronization reveals
a gradual increase between two qubits in the time domain, explaining
the existence of a delay in the superfluorescent pulse of cooperated
radiation from a system-intrinsic point of view. This synchronization
is affected by many factors, among which the symmetry of the system
parameters plays an important part. Tuning the system from a symmetric
setting to an asymmetric setting is accompanied by tuning the transition
rates of the qubits from a synchronous setting to an asynchronous
setting. For the latter, we refer to the scenario where the population
of the left qubit oscillates at a Rabi frequency not synchronous to
that of the right qubit.

However, since two oscillators sharing a common oscillating platform
are able to synchronize after certain time duration according to classical
mechanics, we expect the qubits sharing the cavity resonator would
behave similarly. To precisely describe the transition process from
asynchronous regime to synchronous regime, we extend the quantum synchronization
measure introduced in Ref.~\citep{mari13} for continuous variable
systems to discrete systems. We consider instead the measure of asynchronicity
\begin{equation}
\mathcal{A}\left(t\right)=\Bigl|\det(\rho_{L}\left(t\right)-\rho_{R}\left(t\right))\Bigr|.\label{eq:syn.}
\end{equation}
that compares the difference between two density matrices for two
two-level systems.

\begin{figure}
\includegraphics[bb=10bp 0bp 550bp 495bp,clip,width=8.8cm]{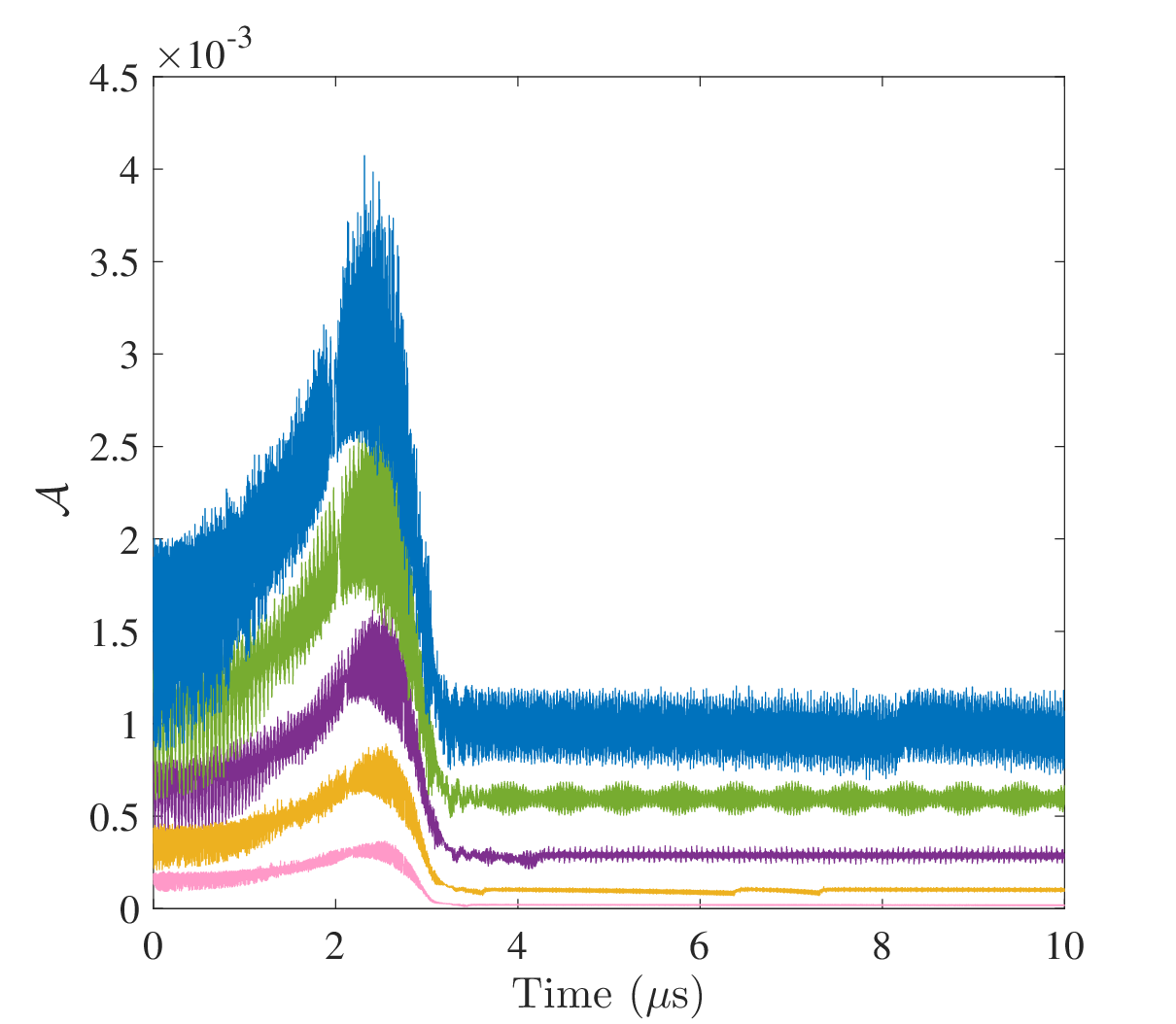}

\caption{Asynchronization $\mathcal{A}$ of two cavity-coupled qubits under
five different coupling strengths: $\eta_{L}/2\pi=\eta_{R}/2\pi=200$MHz
(pink), $300$MHz (yellow), $400$MHz (purple), $500$MHz (green),
and $600$MHz (blue). The qubit level spacings are kept at the asymmetric
setting $\Omega_{L}/2\pi=6.1~\mathrm{GHz}$ and $\Omega_{R}/2\pi=7.1~\mathrm{GHz}$.
\label{fig:syn.}}
\end{figure}

When initiated from the cavity-driven initial state $|\psi\left(0\right)\rangle=3|g,1,g\rangle/\sqrt{10}+|g,2,g\rangle/\sqrt{10}$,
a symmetric setting $\Omega_{L}=\Omega_{R}$ would always lead to
a zero asynchronicity throughout independent of the coupling strengths
$\eta_{L}$ and $\eta_{R}$. When $\Omega_{L}\neq\Omega_{R}$, the
asymmetry leads to coupling-dependent asynchronicity, as shown by
the plots given in Fig.~\ref{fig:syn.} for five settings of coupling
strengths.

No matter the coupling strength, there exists a transition point after
which the asynchronicity remains at a stable value. This transition
point is identical to the transition point shown in Fig.~\ref{fig:3-con.}
(green curves) where the tripartite concurrence reaches a maximal
value. The coincidence verifies our expectation that the synchronization
is maximized when the asynchronicity is minimized. Therefore, synchronization
between two qubits reflects the dynamic identity of the two qubits.

\begin{figure}
\includegraphics[bb=0bp 10bp 680bp 490bp,clip,width=8.8cm]{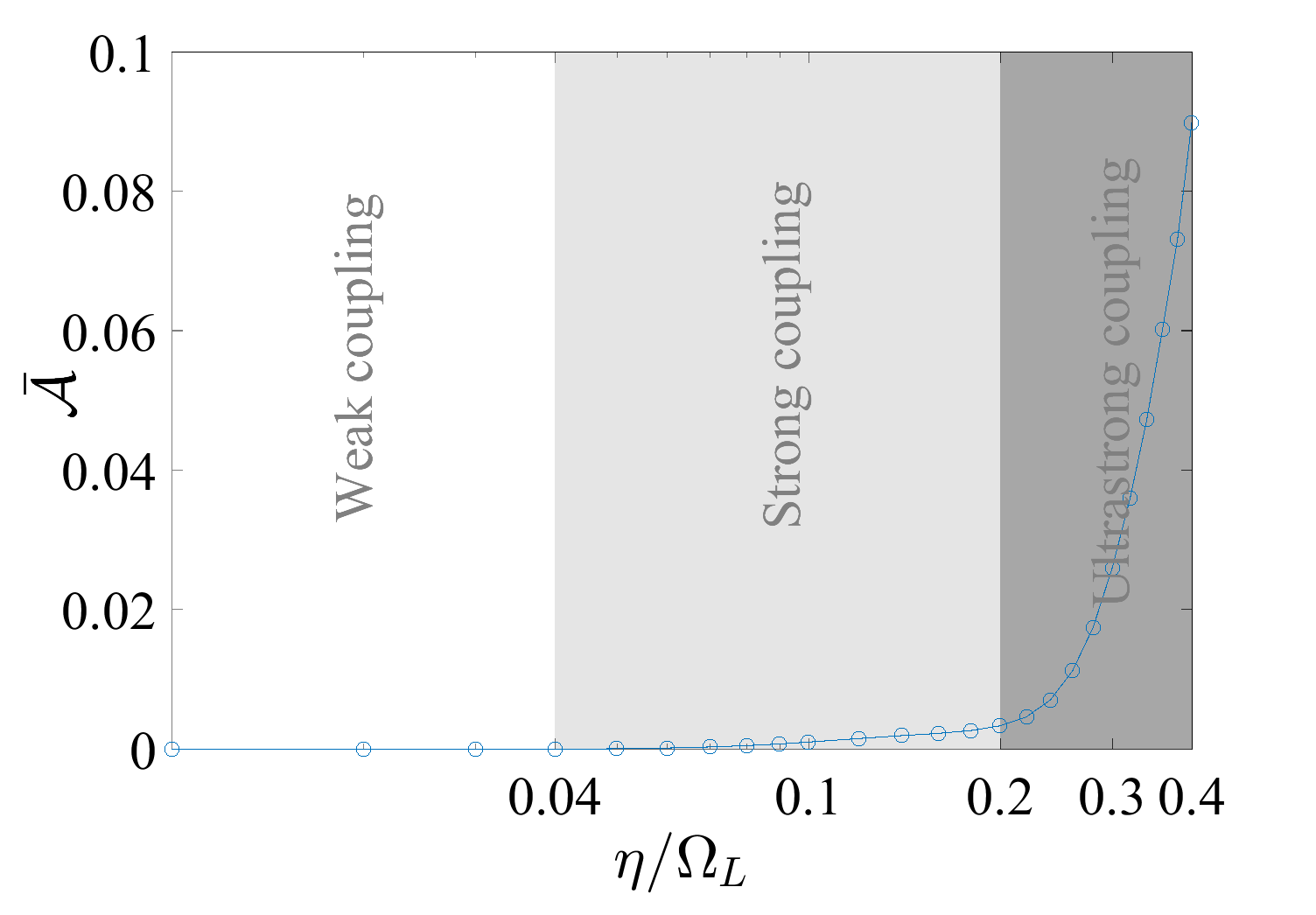}

\caption{The stable value $\bar{\mathcal{A}}$ is plotted as a function of
the coupling strength $\eta=\eta_{L}=\eta_{R}$ in the logarithmic
dimensionless scale of $\eta/\Omega_{L}$, where the qubit level spacings
are kept at the asymmetric setting $\Omega_{L}/2\pi=6.1~\mathrm{GHz}$
and $\Omega_{R}/2\pi=7.1~\mathrm{GHz}$. The circles indicate the
data points from the simulation runnings.\label{fig:syn-eta}}
\end{figure}

Before reaching the stable minimal value, the asynchronicity increases
from a non-zero value for a certain duration, which are spent on the
cooperation by the qubits. When the coupling is sufficiently weak
(below $\eta\approx200$~MHz), the minimal stable value is almost
vanishing (below $10^{-5}$). When the coupling becomes stronger,
the feedback from the cavity mode to each of the qubits becomes adverse
to the synchronizing motion. However, the feedback effect is not linear.
Plotted as a function of the dimensionless coupling strength $\eta/\Omega_{L}$
at $\Omega_{L}/2\pi=6.1~\mathrm{GHz}$ and $\Omega_{R}/2\pi=7.1~\mathrm{GHz}$
in Fig.~\ref{fig:syn-eta}, the stable minimal value \emph{$\bar{\mathcal{A}}$}
of asynchronicity first retains a negligible value in the weak coupling
regime. At about $\eta/\Omega_{L}=0.04$, $\bar{\mathcal{A}}$ starts
to increase slowly until it reaches a turning point at $\eta/\Omega_{L}=0.2$.
The region between $\eta/\Omega_{L}=0.04$ and $\eta/\Omega_{L}=0.2$
can be regarded as a strong coupling regime for synchronization. After
that, $\bar{\mathcal{A}}$ enters the ultra-strong coupling regime
and increases with the coupling again until $\eta/\Omega_{L}\approx0.33$,
where stable minimal value of $\mathcal{A}$ is no longer discernible.

\section{Discussion}

In conclusion, we have studied the synchronization between the two
cavity-coupled qubits using multiple concurrence measures and asynchronicity.
These real-valued measures are computed as functionals of dressed
state vectors that evolve in time as they are driven by an external
field. In all these measures acting as time functions, we obtain consistent
features of transitions from an arbitrary initial system state to
a final synchronized state. The transition in time reveals a synchronization
delay that the qubits use to initiate superfluorescent pulse radiation,
which explains the cooperation origin of the collective effect of
superradiance.

The characteristics of the synchronization process, including the
delay and the value of the stabilized asynchronicity, are highly dependent
on the coupling strengths of the qubits relative to the their level
spacings. They demonstrate from the entanglement perspective the different
behaviors that the circuit QED systems adopt when operating in weak,
strong, and ultra-strong coupling regimes. In general, synchronization
occurs only in the strong- and ultrastrong-coupling regimes whereas
its level of synchronization at the final state increases exponentially
with the qubit-cavity coupling strength.
\begin{acknowledgments}
H. I. thanks the support by FDCT of Macau under grant 065/2016/A2,
University of Macau under MYRG2018-00088-IAPME, and National Natural
Science Foundation of China under grant No.~11404415.
\end{acknowledgments}

\section{Methods}

From the eigen-equation Eq.~(\ref{eq:eigenequation}), one has the
determinant equation

\begin{equation}
\begin{vmatrix}\begin{array}{ccc}
\Delta-E^{\left(n\right)} & \eta_{L} & 0\\
\eta_{L} & \delta_{n}-E^{\left(n\right)} & \eta_{R}\\
0 & \eta_{R} & -\Delta-E^{\left(n\right)}
\end{array}\end{vmatrix}=0
\end{equation}
for each $3\times3$ block of the closed Hamiltonian $H_{0}+H_{\mathrm{int}}$,
where $H_{0}$ contributes the diagonal elements and $H_{\mathrm{int}}$
the off-diagonal elements.

This determinant equation is equivalent to the cubic equation
\begin{equation}
\left(E^{\left(n\right)}\right)^{3}-\delta_{n}\left(E^{\left(n\right)}\right)^{2}-\left(\Delta^{2}+\eta_{L}^{2}+\eta_{R}^{2}\right)E^{\left(n\right)}+\delta_{n}\Delta^{2}-\Delta\left(\eta_{L}^{2}-\eta_{R}^{2}\right)=0,
\end{equation}
whose roots can be derived by absorbing the quadratic term through
the transform $E^{\left(n\right)}=x+\frac{\delta_{n}}{3}$. The transformed
equation becomes $x^{3}+px+q=0$, where 
\begin{align}
p & =-\left(\frac{1}{3}\delta_{n}^{2}+\Delta^{2}+\eta_{L}^{2}+\eta_{R}^{2}\right),\\
q & =-\frac{2}{27}\delta_{n}^{3}-\frac{1}{3}\delta_{n}\left(\eta_{L}^{2}+\eta_{R}^{2}\right)+\frac{2}{3}\delta_{n}\Delta^{2}-\Delta\left(\eta_{L}^{2}-\eta_{R}^{2}\right).
\end{align}

In the close cavity-qubit resonance region $\delta_{n}\thickapprox0$,
the discriminant $D$ is simplified to

\begin{eqnarray}
D & = & \frac{1}{4}\Delta^{2}\left(\eta_{L}-\eta_{R}\right)^{4}-\frac{1}{27}\left(\Delta^{2}+\eta_{L}^{2}+\eta_{R}^{2}\right)^{3}.
\end{eqnarray}
To let the cubic equation admit three non-degenerate real roots, we
consider the range

\begin{eqnarray}
\left(3-2\sqrt{2}\right)\eta_{R}< & \eta_{L} & <\left(3+2\sqrt{2}\right)\eta_{R}
\end{eqnarray}
that makes $D<0$. Applying the Vieta's formula, the roots $x$ can
be found with a parametric angle $\theta_{n}$ given by Eq.~(\ref{eq:theta}).

Using the eigenvalues given in Eq.~(\ref{eq:eigenvalue}) where $k$
indexes the set of states within a cluster of given $n$ and expanding
the eigenvector $\left|u_{k}^{(n)}\right\rangle $ in the bare state
space given by Eq.~(\ref{eq:dressed_state}), we have the column
matrix equation

\begin{equation}
\left[\begin{array}{ccc}
\Delta-E_{k}^{(n)} & \eta_{L} & 0\\
\eta_{L} & \delta_{n}-E_{k}^{(n)} & \eta_{R}\\
0 & \eta_{R} & -\Delta-E_{k}^{(n)}
\end{array}\right]\left[\begin{array}{c}
\alpha_{L,k}^{(n)}\\
\alpha_{C,k}^{(n)}\\
\alpha_{R,k}^{(n)}
\end{array}\right]=0.
\end{equation}
Letting $\alpha_{C,k}^{(n)}$ be the proportional constant, we find
$\alpha_{L,k}^{(n)}=-\eta_{L}\alpha_{C,k}^{(n)}\Bigl/\left(\Delta-E_{k}^{(n)}\right)$
and $\alpha_{R,k}^{(n)}=\eta_{R}\alpha_{C,k}^{(n)}\Bigl/\left(\Delta+E_{k}^{(n)}\right)$.
Then normalizing the coefficients, their expressions are given by
Eqs.~(\ref{eq:alpha_L}) - (\ref{eq:alpha_R}).

\end{document}